\documentclass[aps,11pt]{revtex4}
\usepackage[dvips]{graphicx}

\begin{document}

\title{A Prediction from the Type III See-saw Mechanism}

\author{{\bf S.M. Barr}}
\email{smbarr@bxclu.bartol.udel.edu}
\affiliation{Bartol Research Institute\\
University of Delaware\\
Newark, DE 19716}

\author{{\bf Ilja Dorsner}}
\email{idorsner@ictp.trieste.it} \affiliation{The Abdus Salam
International Centre for Theoretical Physics\\
Strada Costiera 11, 34014 Trieste, Italy}

\begin{abstract}
A simple ansatz that is well-motivated by group-theoretical
considerations is proposed in the context of the type III neutrino
see-saw mechanism. It results in predictions for $m_s/m_b$ and
$m_{\tau}/m_b$ that relates these quantities to the masses and
mixings of neutrinos.
\end{abstract}

\maketitle

Simple unified models based on $SO(10)$ and related groups can
lead to the so-called ``type III see-saw mechanism" for neutrino
masses \cite{type3}. In the most general case the type III
mechanism leads to a light neutrino mass matrix given by the
formula $M_{\nu} = - (M_N H + H^T M_N^T) (u/\Omega)$, where $M_N$
is the Dirac mass matrix of the neutrinos, $H$ is a dimensionless
complex three-by-three matrix and $u/\Omega$ is the ratio of a
weak-scale vacuum expectation value to a GUT-scale vacuum
expectation value (VEV). In a subsequent paper the type III
see-saw mechanism was shown to have certain advantages for
leptogenesis, in particular allowing resonant enhancement without
fine-tuning the form of neutrino mass matrices
\cite{type3leptogen}. In the simplest case, where a minimal set of
Higgs fields breaks $B-L$, one has $H = I$ and the type III
see-saw formula takes the simple form
\begin{equation}
M_{\nu} = - (M_N + M_N^T) \frac{u}{\Omega}.
\end{equation}

The main problem in constructing predictive models of neutrino
masses and mixings with the usual ``type I" see-saw formula
\cite{type1}, $M_{\nu} = - M_N M_R^{-1} M_N^T$, is to relate the
Majorana mass matrix of the right-handed neutrinos $M_R$, with its
six complex parameters, to measurable quantities. There are very
special models, such as the recently much studied ``minimal
$SO(10)$ models", where there is such a relationship
\cite{minimal}. (For an exhaustive list of references on ``minimal
$SO(10)$ models" see~\cite{Aulakh:2005bc}.) And the study of
leptogenesis may tell us something about the structure of $M_R$
(although leptogenesis has only a single data point to work with).
In general, however, the lack of information about $M_R$ is a
problem for the predictivity of type I see-saw models. (The
so-called ``type II see-saw mechanism" \cite{type2} assumes the
existence of $SU(2)_L$-triplet Higgs fields with small VEVs that
couple directly to $\nu_L \nu_L$. About the type II mechanism we
have nothing to say in this paper.)

What makes the simplest version of the type III formula, given in
Eq.~(1), so remarkable and appealing is that it does not involve
the masses of the superheavy right-handed neutrinos at all. As a
consequence, the simplest type III formula opens the possibility
of constructing models of quark and lepton masses that are
extremely predictive. In particular, in models based on $SO(10)$
or other groups that unify an entire family within a single
multiplet, the Dirac mass matrix of the neutrinos $M_N$ is
typically closely related by the grand-unification symmetries to
the mass matrices (also of Dirac type, of course) of the up
quarks, down quarks and charged leptons, which we will denote
respectively as $M_U$, $M_D$, and $M_L$. It is therefore possible
in many models (for examples, see \cite{ab,bpw,ardhs}) to predict
the matrix $M_N$ from a knowledge of the masses and mixings of the
quarks and the masses of the charged leptons. This would allow, if
Eq.~(1) holds, the complete prediction of the mass ratios and
mixing angles of the neutrinos with no free parameters.

In this paper we will not be so ambitious. We have not found so
far a full three-family model that is as predictive as that and
where all the predictions (or ``postdictions") are consistent with
experiment. Rather,
as an illustration of the possibilities of the type III framework,
we will present here a simple ansatz for the heavier two families
that is well motivated by group-theoretical considerations. This
ansatz leads to two interesting predictions that are consistent
with present experimental data. Before presenting the ansatz, we
very briefly review the type III see-saw mechanism and formula.

In models based on $SO(10)$, there are two ways that the
right-handed neutrinos $N^c_i$ ($i = 1,2,3$) can get mass, either
through a renormalizable term such as ${\bf 16}_i {\bf 16}_j
\overline{{\bf 126}}_H$, or through a higher-dimension effective
operator such as ${\bf 16}_i {\bf 16}_j \overline{{\bf 16}}_H
\overline{{\bf 16}}_H/M_{GUT}$. The former allows automatic
conservation of ``matter parity", whereas the latter makes do with
smaller multiplets of Higgs fields. In the latter case, the
effective $d=5$ operator arises most simply from integrating out
three or more $SO(10)$-singlets, which we will denote by ${\bf
1}_a$ or $S_a$, that have the couplings $F_{ia} {\bf 16}_i {\bf
1}_a \overline{{\bf 16}}_H$ and $(M_S)_{ab} {\bf 1}_a {\bf 1}_b$.
If only the Standard-Model-singlet component of the
$\overline{{\bf 16}}_H$ has a non-zero VEV, and we denote it by
$\Omega \sim M_{GUT}$, then one has the familiar ``double see-saw"
mass matrix:
\begin{equation}
{\cal L}_{neutrino} = (\nu_i, N^c_i, S_a) \left(
\begin{array}{ccc}
0 & (M_N)_{ij} & 0 \\ (M_N^T)_{ij} & 0 & F_{ib} \Omega \\
0 & F_{aj} \Omega & (M_S)_{ab} \end{array} \right) \left(
\begin{array}{c} \nu_j \\ N^c_j \\ S_b \end{array} \right).
\end{equation}

\noindent By integrating out the superheavy fields $N^c_i$ and
$S_a$, one obtains $M_{\nu} = - M_N M_R^{-1} M_N^T$, where $M_R =
-(F \Omega) M_S^{-1} (F \Omega)^T$. This is just the type I
see-saw formula, with an effective $M_R$.

Now, if we assume that the $SU(2)_L$-doublet Higgs field contained
in $\overline{{\bf 16}}_H$ also gets a non-zero VEV (and there is
no fundamental reason why it should not), and we denote it by $u$,
then the double see-saw mass matrix takes the form:
\begin{equation}
{\cal L}_{neutrino} = (\nu_i, N^c_i, S_a) \left(
\begin{array}{ccc}
0 & (M_N)_{ij} & F_{ib} u \\ (M_N^T)_{ij} & 0 & F_{ib} \Omega \\
F^T_{aj} u & F^T_{aj} \Omega & (M_S)_{ab} \end{array} \right)
\left(
\begin{array}{c} \nu_j \\ N^c_j \\ S_b \end{array} \right).
\end{equation}

\noindent In this case, it is easy to show that the effective mass
matrix of the light neutrinos takes the form:
\begin{equation}
M_{\nu} = - M_N M_R^{-1} M_N^T - (M_N + M_N^T) \frac{u}{\Omega},
\end{equation}

\noindent where, as before, $M_R = -(F \Omega) M_S^{-1} (F
\Omega)^T$. The first term is the usual type I see-saw
contribution, and the second term is the type III see-saw
contribution. (The origin of the type III term can be simply
understood as follows. One can eliminate the $\nu S$ and $S \nu$
entries in Eq.~(3), i.e. the entries $Fu$ and $F^T u$, by doing a
rotation of the $(\nu_i, N^c_i)$ basis by an angle $\theta \cong
\tan \theta = u/\Omega$. That reduces the matrix in Eq.~(3) to the
same form as Eq.~(2), but with the zeros replaced by terms of the
type III form.) Both the type I and the type III terms in Eq.~(3)
are formally of order $M_W^2/M_{GUT}$. However, since the elements
of $M_N$ are actually small compared to $M_W$ because of small
Yukawa couplings (except perhaps for the third family), and since
$M_N$ comes in quadratically in the type I term but only linearly
in the type III term, one might expect the type III term to
dominate for generic values of the parameters. Moreover, in the
limit that the elements of $M_S$ are small compared to the GUT
scale, the type I contribution becomes small. As was pointed out
in \cite{type3leptogen}, that is a good limit for the purposes of
enhancing leptogenesis. It is therefore plausible that one can
neglect the type I term, and we shall do so.

Now let us turn to the ansatz for the various Dirac mass matrices.
Suppose that these have the form (neglecting the small masses of
the first family)
\begin{equation}
M_U = \left( \begin{array}{ccc} 0 & 0 & 0 \\ 0 & 0 & a \\ 0 & b &
1
\end{array} \right) m_U, \;\;\;\;
M_D = \left( \begin{array}{ccc} 0 & 0 & 0 \\ 0 & 0 & c \\ 0 & d &
1
\end{array} \right) m_D,
\end{equation}
\begin{equation}
M_N = \left( \begin{array}{ccc} 0 & 0 & 0 \\ 0 & 0 & g \\ 0 & h &
1
\end{array} \right) m_U, \;\;\;\;
M_L = \left( \begin{array}{ccc} 0 & 0 & 0 \\ 0 & 0 & e \\ 0 & f &
1
\end{array} \right) m_D,
\end{equation}

\noindent
where the ``texture zero" in the 22 elements can be enforced by an
abelian family symmetry, either discrete or continuous. We will
say more on this later.
And further suppose that the entries satisfy the conditions
\begin{equation}
a + b = g + h, \;\;\;\; c + d = e + f.
\end{equation}

The relations given in Eq.~(7) are not arbitrary, but follow from
group-theory if the elements of the mass matrices come from no
operators except of the following simple types:

(1) ${\bf 16}_i {\bf 16}_j {\bf 10}_H$,

(2) ${\bf 16}_i {\bf 16}_j {\bf 120}_H$,

(3) ${\bf 16}_i {\bf 16}_j {\bf 10}_H {\bf 45}_H/M_{GUT}$.

(4) ${\bf 16}_i {\bf 16}_j {\bf 16}'_H {\bf 16}_H/M_{GUT}$,

\noindent Eq.~(7) is satisfied no matter how many operators there
are of any of these types. Any operator of type (1) gives $a = g$,
$b = h$, $c = e$, and $d = f$, thus satisfying Eq.~(7). Any
operator of type (2) gives contributions that are
flavor-antisymmetric (since the ${\bf 120}$ is in the
antisymmetric product of two spinors). Consequently, it gives $a +
b = 0$, $c + d = 0$, $e + f = 0$, and $g + h = 0$, thus also
satisfying Eq.~(7) in a trivial way.

Any operator of type (3) gives contributions of the form $f_i
f^c_j v_f [\alpha Q(f) + \beta Q(f^c)]$. Here $Q$ is that
generator of $SO(10)$ to which the VEV of the adjoint Higgs field
(${\bf 45}_H$) is proportional; $Q(f)$ is the value of this charge
for the fermion $f$ ($= u, d, \ell^-, \nu$); $v_f = v_u$ or $v_d$
depending on whether $f$ is of the weak-isospin up or down type;
and the coefficients $\alpha$ and $\beta$ depend on the way the
$SO(10)$ indices are contracted in the operator. Thus an operator
of type (3) will give, for instance, $c + d \propto (\alpha +
\beta) (Q(d) + Q(d^c)) v_d$ and $e + f \propto (\alpha + \beta)
(Q(\ell^-) + Q(\ell^+)) v_d$. Since the terms $d_i d^c_j H_d$ and
$\ell^- \ell^+ H_d$ must be invariant under the charge $Q$, it
follows that $Q(d) + Q(d^c) = - Q(H_d) = Q(\ell^-) + Q(\ell^+)$,
and so $c + d = e + f$, satisfying Eq.~(7). In the same way it is
easily seen that $a + b = g + h$.

Finally, consider an operator of type (4). One of the spinor Higgs
fields (say the unprimed one) gets a superlarge VEV that breaks
$SO(10)$ down to $SU(5)$. The effective operator that results is
then of the form $(\alpha {\bf 10}_i \overline{{\bf 5}}_j + \beta
\overline{{\bf 5}}_i {\bf 10}_j) \overline{{\bf 5}}_H$, where the
coefficients depend on the contraction of $SO(10)$ indices in the
original operator. This gives no contribution to $a$, $b$, $g$,
and $h$, and gives contributions to the other paramaters of the
form $c = f$ and $d = e$ (note the transposition between $M_D$ and
$M_L$). Again, such contributions satisfy Eq.~(7).

Simple low-dimension operators that could give contributions {\it
not} satisfying Eq.~(7) are ${\bf 16}_i {\bf 16}_j \overline{{\bf
126}}_H$ (if the $SU(5)$ $\overline{{\bf 45}}$ contained in the
$\overline{{\bf 126}}_H$ got a non-zero VEV), and ${\bf 16}_i {\bf
16}_j \overline{{\bf 16}}'_H \overline{{\bf 16}}_H/M_{GUT}$.

One might ask why we do not include the effects of operators of
even higher dimension, such as ${\bf 16}_i {\bf 16}_j {\bf 10}_H
{\bf 45}_H^n/M_{GUT}^n$, which are not obviously smaller than the
dimension-five operators that we included in our analysis, and
which would not satisfy Eq.~(7) in general. Such operators ought
indeed to be present. However there are reasons that one might
expect them to be small, as we now explain. Consider the operator
${\bf 16}_2 {\bf 16}_3 {\bf 10}_H {\bf 45}_H/ M_{GUT}$, which will
contribute to the 23 and 32 elements in our illustrative model.
Since these elements are somewhat small compared to the 33
elements, either the effective Yukawa couplings in this term are
small or the ratio $\langle {\bf 45}_H \rangle/M_{GUT}$ must be,
or both. This operator can arise from integrating out a pair of
multiplets ${\bf 16}' + \overline{{\bf 16}}'$ that have GUT-scale
mass, as follows. Suppose the terms $a {\bf 16}_3 {\bf 16}' {\bf
10}_H$, $b {\bf 16}_2 \overline{{\bf 16}}' {\bf 45}_H$, and $M{\bf
16}' \overline{{\bf 16}}'$. Integrating out the primed fields
gives an effective operator $a b {\bf 16}_2 {\bf 16}_3 {\bf 10}_H
{\bf 45}_H M^{-1} [1 + |b {\bf 45}_H/M)^2|^{-1/2}$. If $b$ or
$\langle {\bf 45}_H \rangle/M$ are small, then the higher order
operators are highly suppressed. This is not to say that operators
of higher dimension must always be unimportant, but it is a
plausible assumption easily implemented that they can be
neglected.

To return to the texture zero in the 22 elements, it could be
enforced, for example, by a $U(1)$ family symmetry under which the
${\bf 16}_3$, ${\bf 10}_H$ and ${\bf 16}'_H$ are neutral; the
${\bf 16}_2$ has charge $+1$; and the ${\bf 45}_H$, ${\bf 16}_H$,
and ${\bf 120}_H$ have charge $-1$.

Given the simplest type III form (Eq.~(1)), and the ansatz of
Eqs.~(5), (6), and (7), one has
\begin{equation}
M_{\nu} = \left( \begin{array}{ccc} 0 & 0 & 0 \\ 0 & 0 & (a + b) \\
0 & (a + b) & 2 \end{array} \right) \frac{u m_U}{\Omega}.
\end{equation}

\begin{equation}
M_L = \left( \begin{array}{ccc} 0 & 0 & 0 \\ 0 & 0 & e \\
0 & (c+ d - e) & 1 \end{array} \right) m_D,
\end{equation}

\noindent with the quark matrices $M_U$ and $M_D$ given by
Eq.~(5). Consequently, to the extent that we can ignore the first
family, the five parameters $a$, $b$, $c$, $d$, and $e$ determine
the following mass ratios and mixings of the second and third
families: $m_c/m_t$, $m_s/m_b$, $V_{cb}$, $m_{\mu}/m_{\tau}$,
$m_2/m_3$ (the neutrino mass ratio), $U_{\mu 3} \equiv \sin
\theta_{atm}$, and $m_{\tau}/m_b$. Therefore there are two
predictions. (We assume all the parameters are real.)

\begin{figure}[h]
\begin{center}
\includegraphics[width=6.5in]{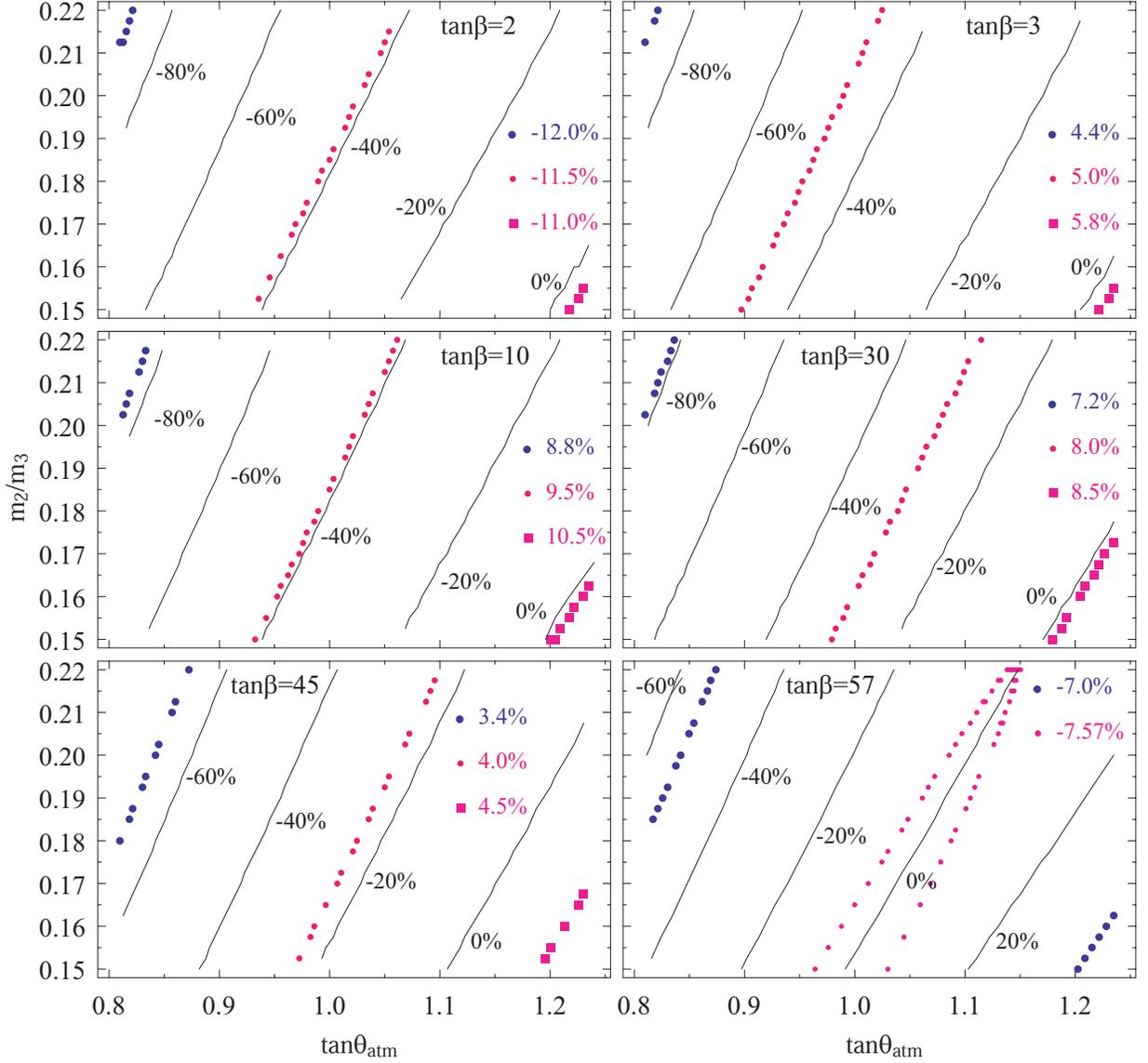}
\end{center}
\caption{\label{figure:1} Plots of constant values of $m_s/m_b$
(solid lines) and $m_{\tau}/m_b$ (dots) in the $\tan
\theta_{atm}$--$m_2/m_3$ plane for $\tan \beta=2,3,10,30,45$, and
$57$. The percents represent the departure of the fitted ratio
from the central value obtained from the RGE evolution. In the
case of the $m_{\tau}/m_b$ ratio the uncertainty in the $b$ and
$\tau$ mass allows this departure to be within $\pm$ 8\% range.}
\end{figure}

What we have done is use the values of the five quantities
$m_c/m_t$, $m_{\mu}/m_{\tau}$, $V_{cb}$, $m_2/m_3$, and
$\theta_{atm}$ to solve for the five parameters $a$, $b$, $c$,
$d$, and $e$. Then we have used the resulting values of those
parameters to ``predict" the values of $m_s/m_b$ and
$m_{\tau}/m_b$ at the GUT scale. For the first three inputs
($m_c/m_t$, $m_{\mu}/m_{\tau}$, and $V_{cb}$), which are fairly
well known, we have taken the central experimental values and run
them up to the GUT scale, assuming low energy supersymmetry. The
running depends significantly on the value of $\tan \beta$, and so
we make a predictions for a particular set of values of $\tan
\beta$ that span the interesting range: 2, 3, 10, 25, 40, and 57.
The other two inputs ($m_2/m_3$ and $\theta_{atm}$) come from
neutrino oscillation experiments (see the reviews
\cite{neutrinos,Fogli:2005cq}) and have rather large error bars.
(For example, $\theta_{atm}=45^\circ \pm 6^\circ$ at $M_Z$.) We
have assumed hierarchical spectrum for neutrino masses with $m_1 <
0.007$\,eV. Under this assumption the RGE evolved values of
$m_2/m_3$ and $\theta_{atm}$ at the GUT scale remain within 3\% of
their low-scale values even for large $\tan \beta$. (For relevant
renormalization group equations
see~\cite{Babu:1993qv,Antusch:2003kp}.) Hence, we drop their
running and allow these two inputs to vary within the
experimentally allowed range and plot our predictions for
$m_s/m_b$ and $m_{\tau}/m_b$ as a function of them in
Fig.~\ref{figure:1}.

We take the experimental values of the quarks from
Ref.~\cite{quarkmasses}, except for $m_s$ for which we use the
results of lattice calculations as given in Ref.~\cite{lattice}
and double the error as suggested in Ref.~\cite{Kim:2004ki}. The
values of the CKM angles and the charged lepton masses are taken
from PDG 2004 \cite{pdg2004}. In presenting our results for
$m_s/m_b$ and $m_{\tau}/m_b$ in Fig.~\ref{figure:1}, we give the
percentage by which the predicted GUT values differ from the
RGE-evolved experimental central values.

In doing the renormalization group running we assume that all the
sparticles have mass of 1 TeV. From $M_Z$ to 1 TeV, the running is
done at one loop, assuming the Standard Model with two Higgs
doublets. From 1 TeV to the GUT scale (taken to be $2 \times
10^{16}$ GeV) we do a two-loop running assuming the MSSM. The
gauge coupling constants are taken from PDG 2004 \cite{pdg2004}.
We present one example of the RGE evolution in
Table~\ref{tab:table1}.
\begin{table}[h]
\caption{\label{tab:table1} Input values at the $M_Z$ scale vs.\
the GUT scale values ($M_{GUT}=2 \times 10^{16}$\,GeV) for $\tan
\beta =45$. In the fermion case we use indicated errors at the
$M_Z$ scale to extract corresponding errors for individual
fermions at the GUT scale. No correlation is taken into account.}
\begin{ruledtabular}
\begin{tabular}{ccc}
              & $\mu=M_Z$ & $\mu=M_{GUT}$ \\
\hline
 $\tan\beta(\mu)$  &     45.00 &     35.36\\
 $v_u(\mu)$ (GeV)  &    174.05 &    117.33\\
 $v_d(\mu)$ (GeV)  &      3.87 &      3.32\\
 $m_u(\mu)$ (MeV)  & $     2.33^{+     0.42}_{    -0.45}$ & $     0.73^{+
0.13}_{    -0.14}$\\
 $m_c(\mu)$ (MeV)  & $     677.^{+      56.}_{     -61.}$ & $     212.^{+
18.}_{     -19.}$\\
 $m_t(\mu)$ (GeV)  & $     181.^{+      13.}_{     -13.}$ & $      93.^{+
42.}_{     -17.}$\\
 $m_d(\mu)$ (MeV)  & $     4.69^{+     0.60}_{    -0.66}$ & $     1.52^{+
0.19}_{    -0.21}$\\
 $m_s(\mu)$ (MeV)  & $     53.8^{+     13.3}_{    -13.3}$ & $     17.4^{+
4.3}_{     -4.3}$\\
 $m_b(\mu)$ (GeV)  & $     3.00^{+     0.11}_{    -0.11}$ & $     1.34^{+
0.08}_{    -0.07}$\\
 $m_e(\mu)$ (MeV)  & $  0.48684727^{+0.00000014}_{-0.00000014}$ & $
0.35620421^{+0.00000010}_{-0.00000010}$\\
 $m_\mu(\mu)$ (MeV)  & $ 102.75140^{+  0.00033}_{ -0.00033}$ & $ 75.20781^{+
0.00024}_{ -0.00024}$\\
 $m_\tau(\mu)$ (GeV)  & $  1.74669^{+  0.00030}_{ -0.00027}$ & $  1.45111^{+
0.00032}_{ -0.00029}$\\
 $|(V_{CKM})_{12}(\mu)|$  &    0.2200 &    0.2199\\
 $|(V_{CKM})_{13}(\mu)|$  &   0.00367 &   0.00300\\
 $|(V_{CKM})_{23}(\mu)|$  &    0.0413 &    0.0337\\
\end{tabular}
\end{ruledtabular}
\end{table}

It should be noted that, even with the assumption that we are
making that the parameters $a$, $b$, $c$, $d$, and $e$ are real,
there are discrete ambiguities of the relative signs of these
parameters. (The overall sign does not matter.) The choice that
gives by far the best fits is $(a,b,c,d,e) = \pm (-,+,+,-,-)$. A
typical set of values is $a \simeq -0.00455$, $b \simeq + 0.9$, $c
\simeq +0.04$, $d \simeq -0.45$, $e \simeq -0.55$, and $f \equiv c
+ d - e \simeq +0.14$.

Note that the value of $a$ is very small. It is this that accounts
for the smallness of $m_c/m_t$. One way that $a$ might be small
naturally (i.e.\ without fine-tuning) using only the set of
operators that satisfy Eq.~(7) is by means of an operator of the
form ${\bf 16}_2 {\bf 16}_3 {\bf 10}_H {\bf 45}_H/M_{GUT}$, where
$\langle {\bf 45}_H \rangle \propto Q = I_{3R} + \epsilon (B-L)$,
$\epsilon \ll 1$, where $I_{3R}$ and $B-L$ are the familiar
$SO(10)$ generators ($I_{3R}$ the diagonal generator of $SU(2)_R$
in the Pati-Salam subgroup, and $B-L$ the baryon minus lepton
number), and where the fields are contracted in such a way that
this generator $Q$ acts on the field ${\bf 16}_2$. (This would
happen, for instance if the effective operator came from
integrating out a ${\bf 16}' + \overline{{\bf 16}}'$ having the
couplings ${\bf 16}_2 \overline{{\bf 16}}' {\bf 45}_H$, ${\bf
16}_3 {\bf 16}' {\bf 10}_H$, and $M {\bf 16}' \overline{{\bf
16}}'$.) This operator would give off-diagonal mass terms for the
up quarks proportional to $Q(u) u^c_3 u_2 + Q(u^c) u^c_2 u_3$.
Since, $I_{3R}(u) = 0$, this would give $a/b = O(\epsilon)$.

The values of $m_s/m_b$ that we predict are satisfyingly close to
the experimental (lattice) results. A couple of things should be
noted in this regard. First, it was long thought that the
Georgi-Jarlskog \cite{gj} relation $(m_s/m_b)_{GUT} = \frac{1}{3}
(m_{\mu}/m_{\tau})_{GUT}$ gave a good fit to the data in SUSY GUT
models. However, the recent lattice calculations have given
results for $m_s$ that are typically only about 0.6 times the
typical values that had been obtained by previous methods. Because
of that, many models which were constructed in the past to give
the Georgi-Jarlskog result, would be off from the current central
experimental/lattice results for $m_s/m_b$ by about $+60\%$. That
compares to the values we are getting, which agree with the
current central value of $m_s/m_b$ for some of the allowed
$(m_2/m_3)-(\theta_{atm})$ parameter space, and are within $20\%$
for a large part of that space.

A second point is that inclusion of the first family is likely to
push up the predicted value of $m_s/m_b$ by about $5\%$. The
reason is that empirically the relation for the Cabbibo angle
$\theta_C \simeq \sqrt{m_d/m_s}$ is known to work very well
\cite{cabbibo}. As is well-known, this formula arises naturally if
the 11 element of the down quark mass matrix vanishes and the 12
and 21 elements are approximately equal \cite{fritzsch}. But then
diagonalizing the 12 block of the down quark mass matrix will push
up the value of the 22 element by a factor of $(1 + |m_d/m_s|)$.

In any event, we see that further improvement in the measurement
of the $\theta_{atm}$, $\delta m^2_{atm}$, $\delta m^2_{sol}$, and
the lattice results for $m_s$, together with an eventual
determination of $\tan \beta$ will allow our simple ansatz, given
in Eq.~(7), to be tested.

\end{document}